\documentclass[12pt]{iopart}

\usepackage{graphicx} 
\usepackage{hyperref} 
\usepackage{cite}

\begin{document}

\title[]{Combined feedback and sympathetic cooling of a mechanical oscillator coupled to ultracold atoms}

\author{Philipp Christoph, Tobias Wagner, Hai Zhong,  Roland Wiesendanger, Klaus Sengstock, Alexander Schwarz and Christoph Becker }

\address{Institut f\"{u}r Laserphysik, Universit\"{a}t Hamburg, Luruper Chaussee 149, 22761 Hamburg, Germany}

\address{Institut f\"{u}r Angewandte Physik, Universit\"{a}t Hamburg, Jungiusstrasse 9-11, 20355 Hamburg, Germany}

\ead{cbecker@physnet.uni-hamburg.de}
\vspace{10pt}
\begin{indented}
\item[]June 2018
\end{indented}

\begin{abstract}
A promising route to novel quantum technologies are hybrid quantum systems, which combine the advantages of several individual quantum systems. 
We have realized a hybrid atomic-mechanical experiment consisting of a  Si$_3$N$_4$ membrane oscillator cryogenically precooled to $500\,\mathrm{mK}$ and optically coupled to a cloud of laser cooled $^{87}$Rb atoms. 
Here, we demonstrate active feedback cooling of the oscillator to a minimum mode occupation of $\bar{n}_\mathrm{m} = 16 \pm 1$ corresponding to a mode temperature of $T_{\mathrm{min}} \approx 200\, \mu \mathrm{K}$. 
Furthermore, we characterize in detail the coupling of the membrane to the atoms by means of sympathetic cooling.
By simultaneously applying both cooling methods we demonstrate the possibility of preparing the oscillator near the motional ground state while it is coupled to the atoms. 
Realistic  modifications of our setup will enable the creation of a ground state hybrid quantum system, which opens the door for coherent quantum state transfer, teleportation and entanglement as well as quantum enhanced sensing applications.
\end{abstract}

\section{Introduction}
Within the growing field of quantum technologies, quantum hybrid systems composed of a mechanical oscillator coupled to an atom-like microscopic quantum object such as spins \cite{Morton2011,Degen2009, Arcizet2011, Kolkowitz2012, Teissier2014, Ovartchaiyapong2014}, semiconductor quantum dots \cite{Kagan2016, Yeo2013, Montinaro2014}, superconducting circuits \cite{Clarke2008, OConnell2010, Fink2016, Palomaki2013, Pirkkalainen2013}  as well as ensembles of cold atoms and single ions \cite{Wang2006, Hunger2010, Camerer2011, Montoya2015, Moeller2017, Bloch2012, Johanning2009} are of special interest. 
The general idea behind these hybrid quantum systems is to pave the way for  technologically valuable multi-tasking capabilities by combining the individual advantages of the different constituents.
These include high scalability and a plethora of different interaction mechanisms on the mechanical side and the realization of qubits with long coherence times and fast high-fidelity quantum gates on the atom side.
Technological prospects range from quantum computation and quantum communication to quantum enhanced sensing.


During the last decade, ground state cooling of a variety of mechanical oscillators via cryogenic or so-called resolved sideband cooling methods has become possible \cite{OConnell2010, Chan2011, Teufel2011, Peterson2016} offering promising prospects for a whole wealth of novel quantum hybrid systems.

We have recently realized a specific hybrid experiment consisting of a cryogenically cooled Si$_3$N$_4$ membrane oscillator coupled to laser cooled $^{87}$Rb atoms via long-range light interactions \cite{Zhong2017}. 
Beyond the possibility to work with laser cooled atoms we routinely prepare Bose-Einstein condensates in a magnetic trap, dipole trap or 3D optical lattice. 
For the purposes presented here we cool the $^{87}$Rb atoms in a magneto-optical trap (MOT) or optical molasses. 
The atom-membrane interaction is generated by an optical lattice, which resonantly couples the atomic motion in the lattice wells to the fundamental mode of the membrane \cite{Vogell2013}. 
We generate the lattice by retro-reflecting a laser beam off the optomechanical system based on a fiber Fabry-P\'{e}rot \textit{membrane-in-the-middle} (MiM) cavity. 
This cavity enhances the hybrid coupling as well as the sensitivity of balanced homodyne detection which we use to measure the motion of the membrane. 
The homodyne signal is also used to prepare the oscillator close to the quantum ground state by active feedback cooling with a dedicated feedback laser beam.

In this paper we report on the first successful combination of active feedback cooling and laser mediated atom-membrane hybrid coupling, which marks an important step towards the realization of a ground state atomic-mechanical hybrid quantum system. 
Specifically, we demonstrate feedback cooling of the membrane oscillator to a minimum mode occupation $\bar{n}_\mathrm{m} = 16 \pm 1$. 
Furthermore, we demonstrate the robustness of the hybrid coupling mechanism in our experiment with respect to all crucial experimentally controllable parameters.
We determine a hybrid cooperativity $C_\mathrm{hybrid} = 150 \pm 10$ through sympathetic cooling of the membrane oscillator by laser cooling the atoms \cite{Joeckel2015}.
Finally, we show that the realization of a quantum hybrid system is within reach by  coupling a feedback cooled membrane oscillator near the quantum ground state to the atomic cloud. Our measurements represent the first successful combination of sympathetic cooling and feedback cooling \cite{Bennett2014}.

The paper is organized as follows. At first, the feedback cooling setup is presented and the experimental results are discussed.
Subsequently, the hybrid coupling scheme and the sympathetic cooling measurements are presented. 
Finally, the principles and results of combined feedback and sympathetic cooling are discussed, followed by concluding remarks on our results and future prospects of our experiment. 

\section{Active feedback cooling}

\begin{figure}
    \includegraphics[width=1\textwidth]{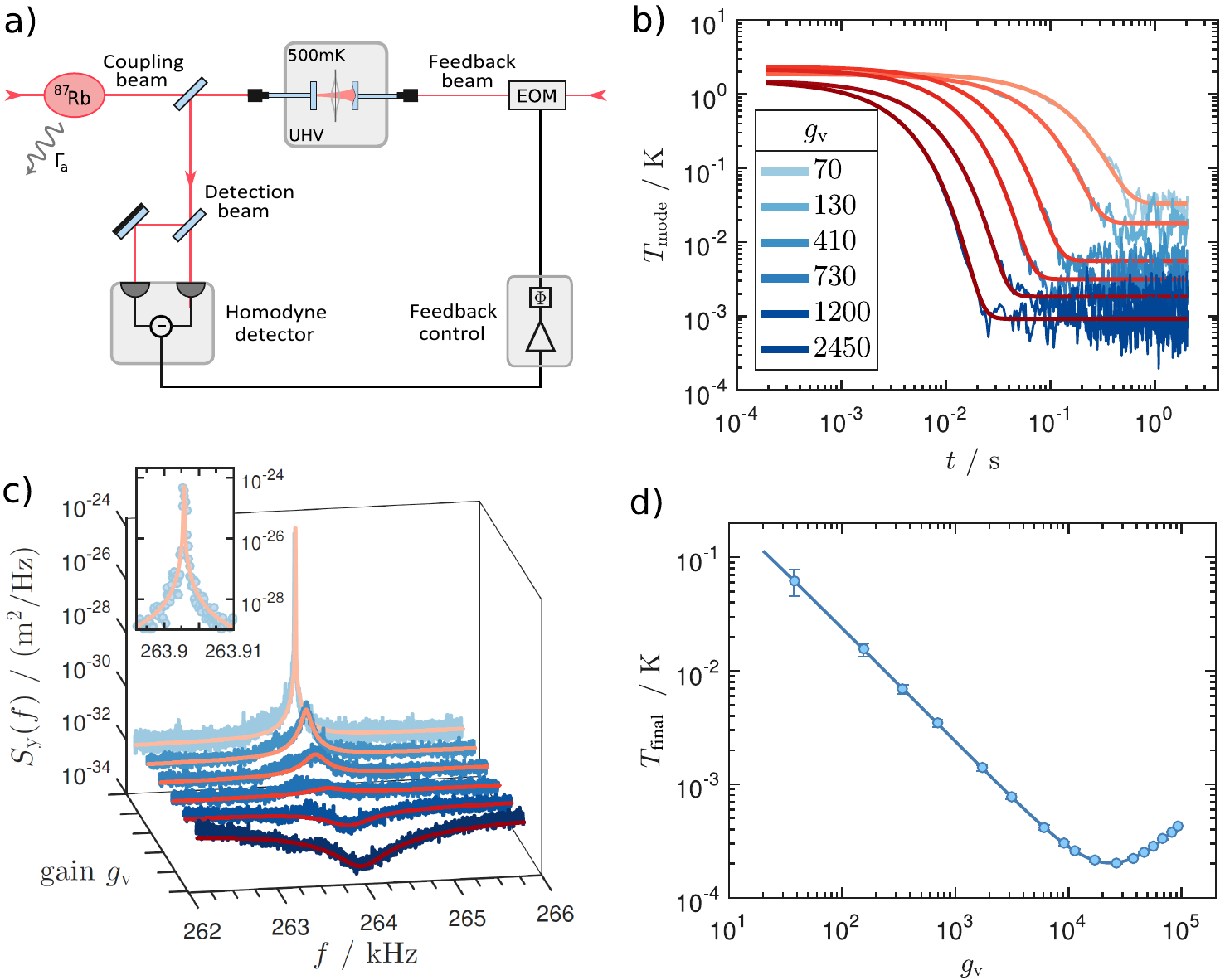}
     \caption{\label{fig:feedback}\textbf{Feedback cooling of the membrane oscillator.} 
     \textbf{a)} 	Scheme of the experimental setup for feedback cooling of a membrane oscillator in a hybrid atomic-mechanical system, as described in the text.
     \textbf{b)} 	Measured zero-span traces of the membrane temperature $T_\mathrm{mode}(t)$ during feedback cooling (blue) for different feedback gains $g_\mathrm{v}$ and fits to 			the data (red) using equation (\ref{eqn:T_mode_t}). Every data set is the average of ten individual measurements.
     \textbf{c)} 	In-loop PSD $S_\mathrm{y}(f)$ measured via Zoom-FFT (blue) for different feedback gains $g_\mathrm{v}$. The fits to the data (red) based on the model in 					\cite{Poggio2007} have $g_\mathrm{v}$ as the only free fitting parameter. Inset: zoom into the peak area for $g_\mathrm{v}=0$.
     \textbf{d)} 	Extracted mode temperatures $T_\mathrm{final}$ as a function  of the fitted $g_\mathrm{v}$ from the spectra in c) and a fit to these mode temperatures according to 			equation (\ref{eqn:T_mode}). We obtain a minimum temperature $T_\mathrm{min} = (203 \pm 15) \, \mathrm{\mu K}$ and a minimum mode occupation $\bar{n}					_\mathrm{m} = 16 \pm 1$.}
\end{figure}

Feedback cooling of mechanical motion \cite{Cohadon1999, Wilson2014} is capable of reaching the motional ground state even in the unresolved sideband regime \cite{Courty2001, Genes2008, Rossi2018}, where optomechanical self-cooling \cite{Chan2011, Teufel2011} is extremely inefficient. 
Strong coupling demands within our hybrid coupling scheme require to operate our fiber cavity \cite{Bick2016, Zhong2017} far in the unresolved sideband regime, which makes feedback cooling the ideal technique to reach the quantum ground state for low-frequency oscillators. We generate feedback by digitally phase-shifting the homodyne signal \cite{Poot2011_Feedback} and feeding it to a fast fiber EOM, which modulates the intensity of a dedicated laser beam. This beam is then coupled into the MiM system and exerts a feedback force on the membrane via radiation pressure, as shown in figure \ref{fig:feedback}\,a. We generate velocity-proportional feedback at gain $g_\mathrm{v}$ by digitally adjusting the effective phase delay to $\Phi_\mathrm{eff} = \pi/2$, which provides optimal cooling \cite{Poot2011,Wilson2014}. By changing the output gain of the digital feedback control we adjust $g_\mathrm{v}$.

When feedback is applied, the membrane quickly reaches a steady state temperature far below its bath temperature $T_\mathrm{bath}$. As shown in figure \ref{fig:feedback}\,b, the temporal behavior of the cooldown can be described well by the model \cite{Pinard2001}
\begin{equation}
T_\mathrm{mode}(t) \ = \ \frac{T_\mathrm{bath}}{1 + g_\mathrm{v}} \, \left( 1+ g_\mathrm{v} \, e^{- \Gamma_\mathrm{m}(1+g_\mathrm{v}   ) t}  \right) \, , \label{eqn:T_mode_t}
\end{equation}
where $\Gamma_\mathrm{m} = 2\pi \times 24.5 \, \mathrm{mHz}$ is the natural linewidth of the ground mode $\omega_\mathrm{m} = 2\pi \times 264 \, \mathrm{kHz}$ of our membrane oscillator \cite{Zhong2017}. After a few cooldown times $t_\mathrm{cool} = (\Gamma_\mathrm{m} g_\mathrm{v})^{-1}$ the final temperature $T_\mathrm{final}$ is reached. $T_\mathrm{final}$ just depends on the feedback gain $g_\mathrm{v}$, as long as the measured signal is well above our detection noise floor $S_\mathrm{x_n} = 7.4 \cdot 10^{-33} \, \mathrm{m^2/Hz}$ \cite{Poggio2007}: 
\begin{equation}
T_\mathrm{final} \ = \ \frac{T_\mathrm{bath}}{1  + g_\mathrm{v}} \, + \left[ \frac{m \omega_\mathrm{m}^3}{4 k_\mathrm{B} Q_\mathrm{m} } \, \frac{g_\mathrm{v}^2}{1 + g_\mathrm{v}} \right] S_\mathrm{x_n}   \, . \label{eqn:T_mode}
\end{equation}
Here, $m = 76 \, \mathrm{ng}$ denotes the effective mass of the membrane and $Q_\mathrm{m} = \omega_\mathrm{m}/\Gamma_\mathrm{m}$ its quality factor. 
At very large feedback gains, we observe strong noise-squashing \cite{Aspelmeyer2014} of our in-loop measured power spectral density (PSD) $S_\mathrm{y}(f)$, as shown in figure \ref{fig:feedback}\,c. 
We obtain the mode temperature by fitting $S_\mathrm{y}(f)$ and subsequently calculating the real out-of-loop PSD $S_\mathrm{x}(f)$ and finally integrating these spectra \cite{Poot2011}. 
Analytically, this is equivalent to equation (\ref{eqn:T_mode}) using $g_\mathrm{v}$ from the spectral fits \cite{Poggio2007}. 
Figure \ref{fig:feedback}\,d shows a fit to these final temperatures yielding a minimum temperature of $T_\mathrm{min} = (203 \pm 15) \, \mathrm{\mu K}$ which corresponds to a minimum mode occupation of $\bar{n}_\mathrm{m} = 16 \pm 1$. 
The error is mainly given by the systematic calibration error of our homodyne detection, which we obtained by sweeps of the cryostat temperature \cite{Wilson2014} and which agrees well with our optomechanical calibration method presented in \cite{Zhong2017}.

Ground state feedback cooling is possible if $S_\mathrm{x_n} < 4 x_\mathrm{zp}^2 / (n_\mathrm{th} \Gamma_\mathrm{m})$ \cite{Wilson2014}, where $x_\mathrm{zp} = [\hbar / (2 m \omega_\mathrm{m})]^{1/2}$ denotes the zero-point motion of the oscillator and $n_\mathrm{th} \approx k_\mathrm{B} T_\mathrm{bath} / (\hbar \omega_\mathrm{m})$ its thermal phonon occupation. 
We expect ground state cooling in our setup for a ten times larger quality factor $Q_\mathrm{m}$ and a ten times lower mass of the oscillator, which can easily be met by using new types of mechanical oscillators \cite{Norte2016, Reinhardt2016}.
Reducing $T_\mathrm{bath}$ or improving the detection noise floor $S_\mathrm{x_n}$ will even relax this condition.

\section{Sympathetic cooling}


The coupling mechanism for sympathetic cooling can be easily understood on the basis of two coupled harmonic oscillators, one being represented by the membrane oscillator $\omega_\mathrm{m}$, the other by atoms oscillating in the individual potential wells of a deep optical lattice with frequency $\omega_\mathrm{a}$ \cite{Vogell2013}.
Atoms are loaded in a 1D optical lattice whose retroreflection mirror is replaced by a membrane oscillator inside an optical cavity. The vibrating motion of the membrane slightly changes the cavity$'$s resonance condition leading to a phase modulation of the back-reflected light. As a consequence, the lattice wells in which the atoms are trapped are periodically displaced, which can excite the atoms to higher levels in the corresponding harmonic oscillator potential of each individual lattice well. In turn, a displacement of the atoms in the optical lattice potential wells leads to a redistribution of photons between the two counterpropagating lattice beams related to the optical dipole force \cite{Raithel1998} and modulates the light intensity and consequently the corresponding radiation pressure acting on the membrane.  In this way, a bidirectional resonant coupling is realized if $\omega_\mathrm{a} = \omega_\mathrm{m}$ as described in detail in \cite{Vogell2013}.

The effective coupling Hamiltonian $\hat{H}_\mathrm{eff} \sim  g_\mathrm{N} \left( \hat{a}_\mathrm{at}^\dagger  \hat{a}_\mathrm{m} +  \hat{a}_\mathrm{m}^\dagger \hat{a}_\mathrm{at} \right) $ is of beam-splitter type and allows to exchange energy between the two systems on the single phonon level at rate $g_\mathrm{N}$. If the atoms are laser cooled, sympathetic cooling of the oscillator can be realized with the potential for ground state cooling even for low-frequency oscillators. 

\begin{figure}
    \includegraphics[width=1\textwidth]{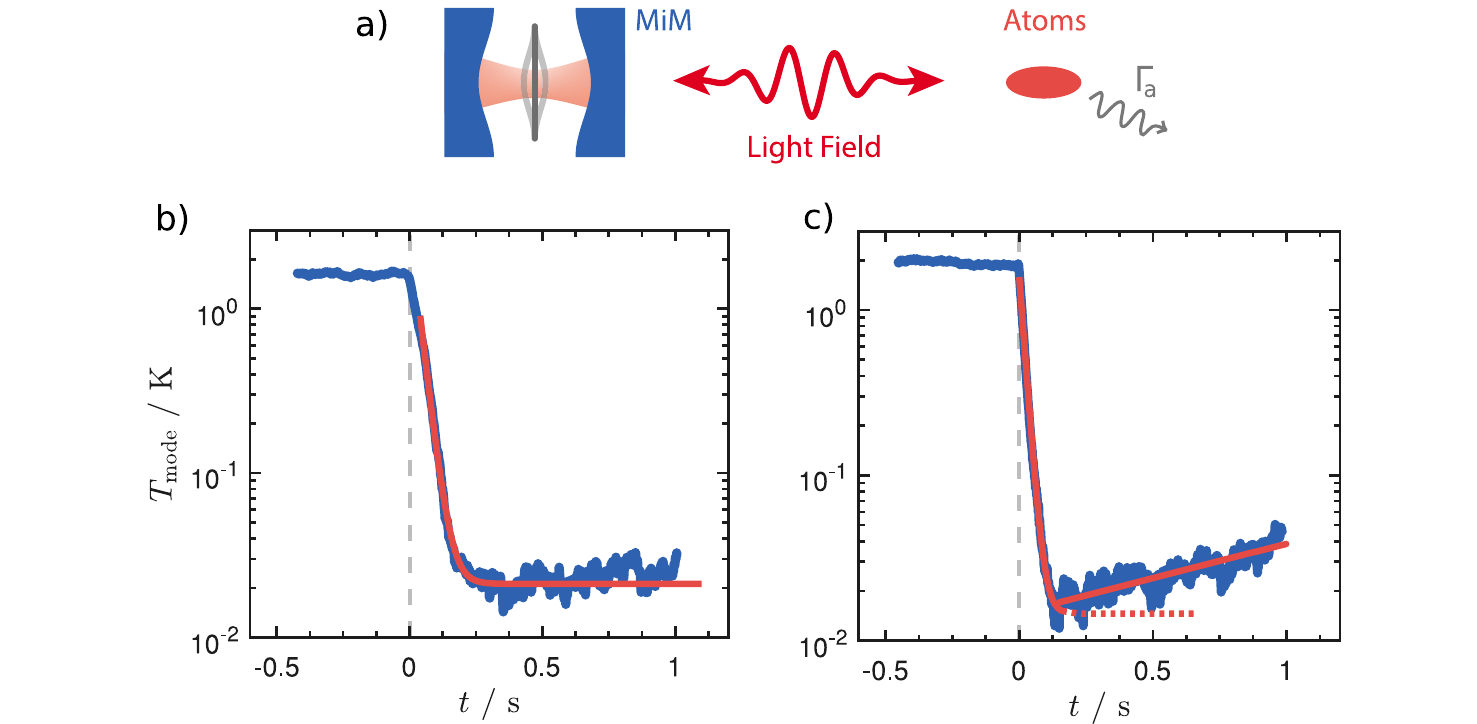}
     \caption{\label{fig:sympathetic_timetrace}\textbf{Sympathetic cooling in time-domain.} 
     \textbf{a)} 	Laser cooling the atoms at rate $\Gamma_\mathrm{a}$ leads to sympathetic cooling of the oscillator in the MiM system. 
     \textbf{b)} 	Zero-span trace (blue) of the membrane temperature $T_\mathrm{mode}$, averaged over 30 experimental runs and fit (red) to the data for sympathetic MOT cooling, 			as described in the text. Parameters: blue lattice detuning $\Delta_\mathrm{lat}^{2,3} = 2\pi \times 0.507 \, \mathrm{GHz}$, $P_\mathrm{lat} = 0.5 \, \mathrm{mW}$, $			\omega_\mathrm{a} = 2.5 \, \omega_\mathrm{m}$ (calibrated). 
     \textbf{c)} 	Same as b) but using optical molasses cooling. Parameters: red lattice detuning $\Delta_\mathrm{lat}^{2,1} = -2\pi \times 1.35 \, \mathrm{GHz}$, $P_\mathrm{lat} = 				0.56 \, \mathrm{mW}$, $\omega_\mathrm{a} = 1.48 \, \omega_\mathrm{m}$ (calibrated).}
\end{figure}

The sympathetic cooling rate $\Gamma_\mathrm{sym}$ depends on many different parameters. Besides the MiM properties, these are mainly the number of atoms $N$, the coupling lattice parameters and the laser cooling rate $\Gamma_\mathrm{a}$ \cite{Joeckel2015} ($ r_\mathrm{m} $ denotes the membrane's reflectivity and $\mathcal{F}$ the cavity finesse):

\begin{equation}
\Gamma_\mathrm{sym}[N, \omega_\mathrm{a}] \ =  \ \frac{g_\mathrm{N}^2 \Gamma_\mathrm{a}}{\left( \omega_\mathrm{a} - \omega_\mathrm{m}  \right)^2 + \left( \Gamma_\mathrm{a}/2 \right)^2   } \ ,   \  g_\mathrm{N} \,  = \, \vert r_\mathrm{m} \vert^2 \omega_\mathrm{a} \sqrt{\frac{N m_\mathrm{a} \omega_\mathrm{a}}{m \omega_\mathrm{m}  }  } \frac{2 \mathcal{F}}{\pi} \, . \label{eqn:Gamma_sym}
\end{equation}
We performed extensive characterization measurements of $\Gamma_\mathrm{sym}$ which will be presented in the following. 
Through these measurements we also determine the hybrid cooperativity $C_\mathrm{hybrid} = 4 g_\mathrm{N}^2 / ( \Gamma_\mathrm{a} \Gamma_\mathrm{m}) =  \Gamma_\mathrm{sym} / \Gamma_\mathrm{m}$, which summarizes the system's capabilities to operate in the strong coupling regime.

We start by preparing samples of laser cooled atoms. Subsequently, we quickly ramp up the coupling lattice within $1 \, \mathrm{ms}$ to a variable final value of $\omega_\mathrm{a}$ that we calibrate separately using Kapitza-Dirac diffraction of a Bose-Einstein condensate \cite{Ovchinnikov1999}. 
The optical lattice is near-detuned with $- 2 \, \mathrm{GHz} < \Delta_\mathrm{lat}/2 \pi <  2 \, \mathrm{GHz}$ for the measurements presented here. 
During this sequence we continuously monitor the temperature of the membrane $T_\mathrm{mode}$ as a function of time. 
Exemplary time traces of $T_\mathrm{mode}$ are shown in figure \ref{fig:sympathetic_timetrace}\,b for MOT cooling and in \ref{fig:sympathetic_timetrace}\,c for optical molasses cooling. 
The sympathetic cooling leads to minimum mode temperatures $T_\mathrm{min} \approx 20 \, \mathrm{mK}$ within approximately $100 \, \mathrm{ms}$. We obtain the corresponding $\Gamma_\mathrm{sym}$ from $T_\mathrm{min}$ using:

\begin{equation}
T_\mathrm{min}  \  =   \ T_\mathrm{bath} \, (1 + \Gamma_\mathrm{sym} / \Gamma_\mathrm{m})^{-1}  \, . \label{eqn:T_min}
\end{equation}
As the time dependence of sympathetic cooling can be described in the same formalism as feedback cooling \cite{Bennett2014}, the cooldown data can be fitted well with equation (\ref{eqn:T_mode_t}). 

For MOT cooling as shown in figure \ref{fig:sympathetic_timetrace}\,b, we observe that $T_\mathrm{mode}$ reaches a quasi steady state at $T_\mathrm{min}$ before the decay of the MOT leads to an increase of $T_\mathrm{min}$ on the time scale of several seconds (barely visible in this figure). We have optimized our MOT  for maximum cooling performance, which is achieved when quickly ramping the laser cooling parameters to new values before switching on the coupling lattice in the following way. We simultaneously ramp the MOT detuning $\Delta_\mathrm{MOT}$, intensity $I_\mathrm{MOT}$ and magnetic field gradient $B'$ from $\Delta_\mathrm{MOT} = 2.9 \,  \Gamma_\mathrm{D_2}$ to $6.2 \,  \Gamma_\mathrm{D_2}$, $I_\mathrm{MOT} = 50 \, \mathrm{mW/cm^2}$ to $4 \, \mathrm{mW/cm^2}$ and $B' = 5 \, \mathrm{G/cm}$ to $45 \, \mathrm{G/cm}$, correspondingly. We have checked with independent OD measurements, that this produces particularly dense atomic samples.

For molasses cooling we observe that $T_\mathrm{min}$ increases exponentially with time as seen in figure \ref{fig:sympathetic_timetrace}\,c, 
This is  due to atomic diffusion out of the lattice volume, which can be fitted well with an exponentially decreasing atom number $N$ in $\Gamma_\mathrm{sym}$ (see equation (\ref{eqn:Gamma_sym})) using equation (\ref{eqn:T_min}).
We iteratively optimize intensity and detuning of the optical molasses to find the maximal $\Gamma_\mathrm{sym}$ and observe the expected behavior \cite{Joeckel2015}. 

All sympathetic cooling measurements presented in the following were performed with these optimized MOT and molasses parameters.
It is important to note that once the high density MOT or the optical molasses is generated, the requirements on timing-related lattice parameters like ramping speed or wait time before the ramp up  are very relaxed, which makes sympathetic cooling in our system extremely robust.

\begin{figure}
    \includegraphics[width=1\textwidth]{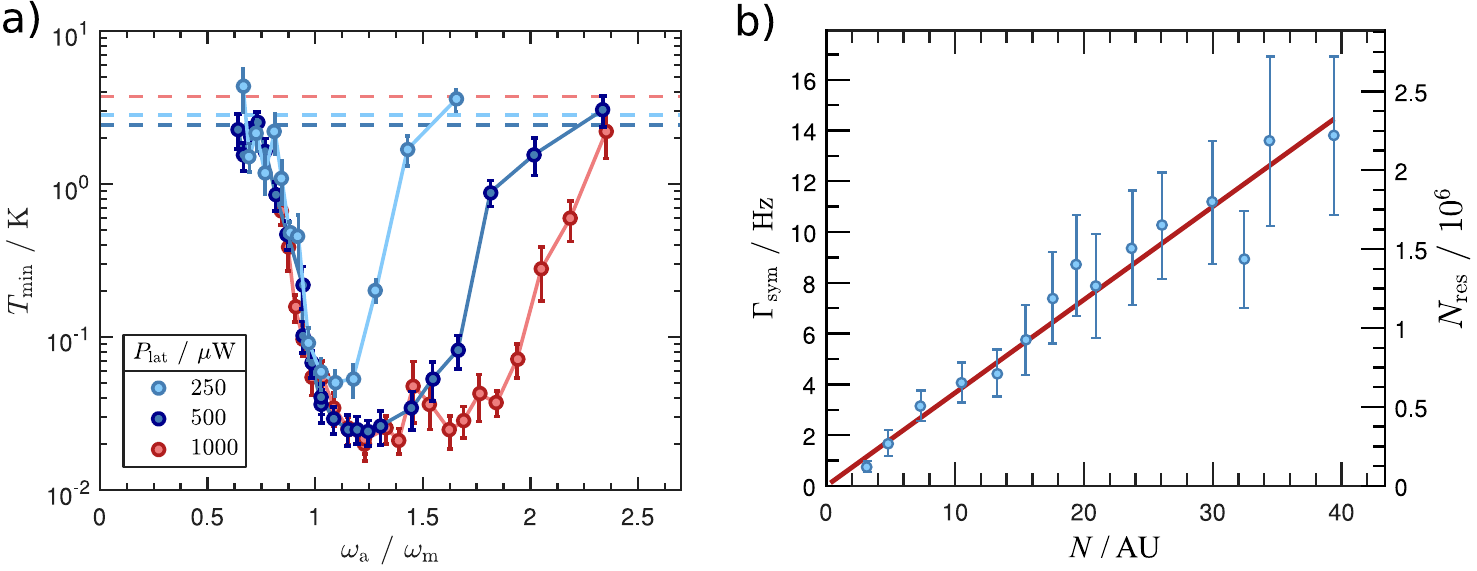}
     \caption{\label{fig:sympathetic_validation}\textbf{Robustness of sympathetic cooling.} 
     \textbf{a)} 	$T_\mathrm{min}(\omega_\mathrm{a})$ for molasses cooling (dots) with three different $P_\mathrm{lat}$ and $T_\mathrm{bath}$ indicated by dashed lines (average of 			$T_\mathrm{mode}$ for $t<0$, see figure \ref{fig:sympathetic_timetrace}). The solid lines are guides to the eye. 
     \textbf{b)} 	Measured $\Gamma_\mathrm{sym}$ for MOT cooling as a function of the atom number $N$ and calculated $N_\mathrm{res}$ as described in the text. Parameters: 			blue lattice detuning $\Delta_\mathrm{lat}^{2,3} = 2\pi \times 0.48 \, \mathrm{GHz}$, $P_\mathrm{lat} = 0.35 \, \mathrm{mW}$, $\omega_\mathrm{a} = 2.1 \, 					\omega_\mathrm{m}$ (calibrated). The red line is a linear fit to the data.}
\end{figure}

In order to characterize the coupling mechanism further, we varied $\omega_\mathrm{a}$ around $\omega_\mathrm{m}$ by changing the lattice detuning at a given lattice power $P_\mathrm{lat}$. 
The resulting minimum temperatures $T_\mathrm{min}$ were measured as described in figure \ref{fig:sympathetic_timetrace} and the result is shown in figure \ref{fig:sympathetic_validation}\,a.
As expected the lowest temperature $T_\mathrm{min}$ is achieved for $\omega_\mathrm{a} > \omega_\mathrm{m}$ \cite{Joeckel2015}.
Moreover the measurements show that the resonance condition of the sympathetic cooling mechanism around $\omega_\mathrm{m}$ is not very critical and allows for robust operation.

Unlike the predicted monotonous increase of $\Gamma_\mathrm{sym}$ for larger $\omega_\mathrm{a}$, we observe that the sympathetic cooling rate decreases if $\omega_\mathrm{a}$ is increased beyond a certain point.
This decrease in cooling rate is more pronounced for small lattice detunings as seen in figure \ref{fig:sympathetic_validation}\,a.
We explain this by increased light scattering for near-detuned optical lattice light.
Another important parameter in the hybrid coupling scheme is the number of atoms in the lattice volume $N$, which was altered systematically by changing the MOT loading time and calculating $\Gamma_\mathrm{sym}$ from the measured $T_\mathrm{min}$ according to equation (\ref{eqn:T_min}).
According to equation (\ref{eqn:Gamma_sym}) the sympathetic cooling rate depends linearly on $N$, which we confirm in figure \ref{fig:sympathetic_validation}\,b.
We observe this behavior for molasses cooling and for MOT cooling with a blue detuned lattice.
The qualitative value of $N$ was obtained by measuring the optical density along the lattice beam using a mode matched, weak probe beam.
We obtain $N_\mathrm{res}$ from the fitted $\Gamma_\mathrm{sym}$ and plot it as the second y-axis of figure \ref{fig:sympathetic_validation}\,b.

\begin{figure}
    \includegraphics[width=1\textwidth]{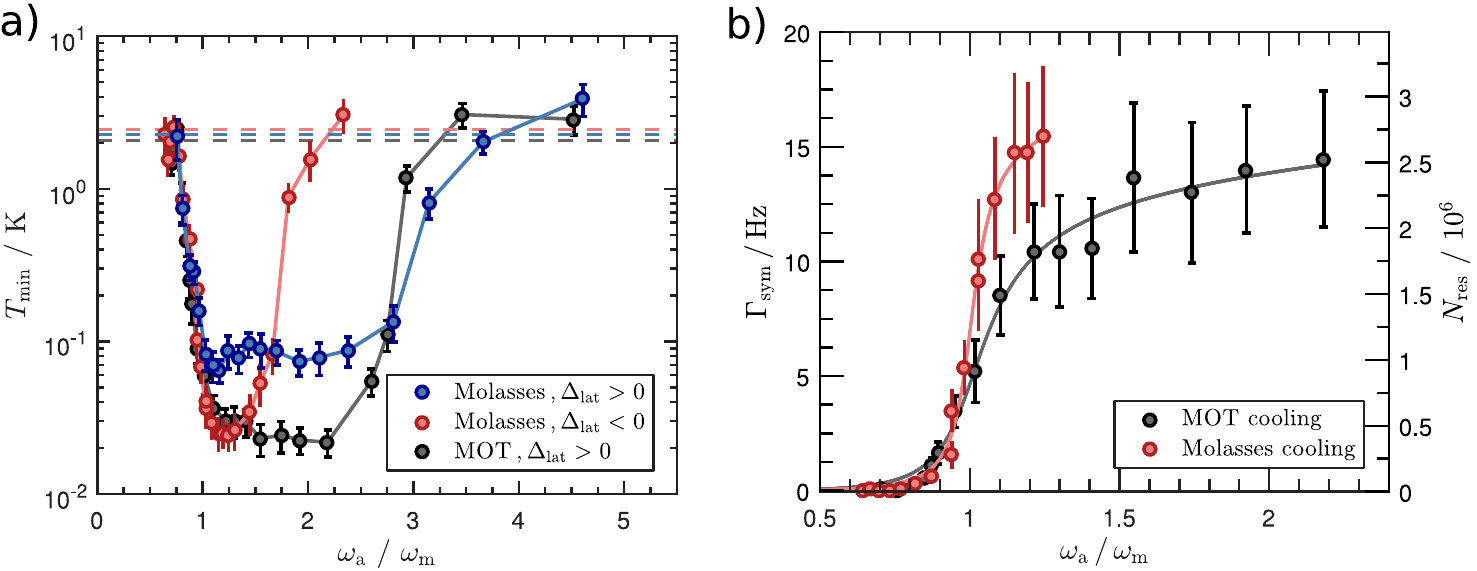}
     \caption{\label{fig:sympathetic_methods}\textbf{Optimized sympathetic cooling results.}
     \textbf{a)} 	$T_\mathrm{min}(\omega_\mathrm{a})$ for molasses cooling with a blue detuned lattice (blue dots) and red detuning (red dots) and for MOT cooling (black dots) with a 			blue detuned lattice ($P_\mathrm{lat} = 0.5 \, \mathrm{mW}$ in all three cases). The solid lines are guides to the eye. $T_\mathrm{min}$ and $T_\mathrm{bath}$ 				(dashed lines) obtained as in figure \ref{fig:sympathetic_validation}. The x-axis was calibrated by the fit in b). 
     \textbf{b)} 	Extracted $\Gamma_\mathrm{sym}$ and $N_\mathrm{res}$ for the best cooling curves in a). Solid lines are fitted, ensemble-integrated cooling rates as described in 			the text}
\end{figure}

All tested laser cooling methods with experimental parameters individually optimized are summarized in figure \ref{fig:sympathetic_methods}. We observe the best sympathetic cooling down to $T_\mathrm{min} \approx 20 \, \mathrm{mK}$ with optical molasses cooling using a red detuned lattice ($\Delta_\mathrm{lat} < 0$) and for a MOT cooling with a blue detuned lattice ($\Delta_\mathrm{lat} > 0$). As these two measurements were performed with exactly the same lattice configuration (except for the sign of $\Delta_\mathrm{lat}$), the faster increase of $T_\mathrm{min}(\omega_\mathrm{a})$ for molasses cooling can not be explained only by the light scattering mentioned above. Probably, the additional parasitic effect is caused by the known instability of the hybrid coupling mechanism \cite{Aline2017}, which is predicted to be larger for red lattice detunings \cite{Asboth2007, Asboth2008}.
This assumption is in agreement with our observation that sympathetic cooling is drastically reduced and turned into heating for MOT cooling with a red detuned lattice.
Sympathetic molasses cooling is less efficient for a blue detuned lattice as the repulsive potential of the lattice beam leads to a reduction of the number of atoms $N$ in the coupling volume.
For the two best cooling curves in figure \ref{fig:sympathetic_methods}\,a we extracted the corresponding $\Gamma_\mathrm{sym}$ using equation (\ref{eqn:T_min}), as shown in figure \ref{fig:sympathetic_methods}\,b.
We find that our measured $\Gamma_\mathrm{sym}$ agrees very well with the theoretically expected cooling rate based on an ensemble-integrated model \cite{Joeckel2015}.
Furthermore, the fit allows to extract the value of the laser cooling rate $\Gamma_\mathrm{a} = (0.11 \pm 0.03) \, \omega_\mathrm{m}$ for molasses cooling and $\Gamma_\mathrm{a} = (0.24 \pm 0.07) \, \omega_\mathrm{m}$ for MOT cooling.
The fit can also be used to precisely calibrate the value of $\omega_\mathrm{a}$, which in this case was also used to calibrate the x-axis of figure \ref{fig:sympathetic_methods}\,a. $N_\mathrm{res}$ was calculated from $\Gamma_\mathrm{sym}$ as described above.

The maximum sympathetic cooling rate we could achieve was measured independently with molasses cooling using the same parameters as quoted in figure \ref{fig:sympathetic_methods}\,b at the optimal point $\omega_\mathrm{a} = 1.25 \, \omega_\mathrm{m}$. 
For this, we further increased the atom number in the optical molasses leading to a maximum cooling rate of $\Gamma_\mathrm{sym} = 23.3 \pm 1.4 \, \mathrm{Hz}$. 
This corresponds to a hybrid cooperativity $C_\mathrm{hybrid} = 150 \pm 10$. As $C_\mathrm{hybrid} \ll n_\mathrm{th} \approx 2 \cdot 10^5$ in our current setup, we operate far outside the strong coupling regime \cite{Vogell2013}.
However, as discussed in more detail in the conclusion realistic improvements on our optomechanical MiM system will enable us to enter the strong coupling regime.

\section{Feedback-assisted sympathetic cooling}
%
The realization of a strongly coupled hybrid quantum system in our setup  demands for cooling and coupling at the same time.
In the following we study the prospects of reaching the strongly coupled regime by experimentally characterizing simultaneous sympathetic and feedback cooling and comparing the results to a  classical model for the measured in-loop PSD $S^\mathrm{sf}_\mathrm{y}(\omega)$ of combined sympathetic and feedback cooling, which we use to fit the measured spectra. 

Consider the equation of motion for the membrane in frequency-domain 

\begin{equation}
x \ = \ \chi_\mathrm{m} \left[ F_\mathrm{th} + F_\mathrm{fb} + F_\mathrm{sym}  \right] \, . \label{eqn:x}
\end{equation}
Here, $\chi_\mathrm{m}$ is the mechanical susceptibility of the membrane, $F_\mathrm{th}$ denotes the random thermal force acting on the membrane due to its coupling to the environment, $F_\mathrm{fb}$ is the feedback force and $F_\mathrm{sym}$ the sympathetic cooling force given by the hybrid coupling mechanism. Note that all frequency dependencies were omitted for clarity and that all quantum noise contributions are neglected. The sympathetic cooling force can be approximated by $F_\mathrm{sym} = - \chi_\mathrm{sym}^{-1} x \approx i m \omega \Gamma_\mathrm{sym} x$ \cite{Joeckel2015}, while the feedback force also  includes the detector noise contribution $x_\mathrm{n}$: $F_\mathrm{fb} = - \chi_\mathrm{fb}^{-1}(x + x_\mathrm{n})$ \cite{Poot2011, Wilson2014}. In the case of velocity-proportional feedback, the feedback transfer function is given by $\chi_\mathrm{fb}^{-1} = - i m \omega \Gamma_\mathrm{m} g_\mathrm{v}$. Inserting this into the equation of motion (\ref{eqn:x}) yields:

\begin{eqnarray}
\fl \qquad \qquad  \left( \chi_\mathrm{m}^{-1} + \chi_\mathrm{fb}^{-1} + \chi_\mathrm{sym}^{-1}   \right) \,  x \ &\equiv  \  \chi_\mathrm{eff,sf}^{-1} \, x \ = \ F_\mathrm{th} - \chi_\mathrm{fb}^{-1} \, x_\mathrm{n} \label{eqn:Symp_Chi_eff_sf_x} \\
 \fl \Leftrightarrow  \left( \chi_\mathrm{m}^{-1} + \chi_\mathrm{fb}^{-1} + \chi_\mathrm{sym}^{-1}   \right) \, \left( x +  x_\mathrm{n} \right) \ & \equiv  \  \chi_\mathrm{eff,sf}^{-1} \, y \ = \ F_\mathrm{th} + \left( \chi_\mathrm{m}^{-1} + \chi_\mathrm{sym}^{-1} \right) \, x_\mathrm{n} \, . \label{eqn:Symp_Chi_eff_sf_y} 
 \\[-0.4cm]
\fl & \qquad \qquad \qquad \qquad   \ \   \underbrace{\qquad \qquad \quad \ }_{\chi_\mathrm{eff,s}^{-1}} \nonumber
\end{eqnarray}
Here, we have introduced the effective susceptibilities for combined cooling $\chi_\mathrm{eff,sf} = \chi_\mathrm{m} \left[ \Gamma_\mathrm{m}' = \Gamma_\mathrm{m}(1 + g_\mathrm{v} + g_\mathrm{s}) \right]$ and for sympathetic cooling $\chi_\mathrm{eff,s} = \chi_\mathrm{m} \left[\Gamma_\mathrm{m}' = \Gamma_\mathrm{m}(1 +  g_\mathrm{s}) \right]$ with the effective sympathetic cooling gain $g_\mathrm{s} = \Gamma_\mathrm{sym}/\Gamma_\mathrm{m} $. From equations (\ref{eqn:Symp_Chi_eff_sf_x}) and (\ref{eqn:Symp_Chi_eff_sf_y}) we deduce the measured in-loop PSD $S^\mathrm{sf}_\mathrm{y}(\omega)$ and the real out-of-loop PSD $S^\mathrm{sf}_\mathrm{x}(\omega)$:

\begin{eqnarray}
S_\mathrm{x}^\mathrm{sf}(\omega) \ &= \ \langle \,  \vert x(\omega) \vert^2 \rangle \ = \ \vert \chi_\mathrm{eff,sf} \vert^2 \, \left[  S_\mathrm{F_{th}}(\omega) + \vert \chi_\mathrm{fb}\vert^{-2} S_\mathrm{x_{n}}(\omega) \right] \label{eqn:Symp_S_x} \\
S_\mathrm{y}^\mathrm{sf}(\omega) \ &= \ \langle \, \vert y(\omega) \vert^2 \rangle \ = \ \vert \chi_\mathrm{eff,sf} \vert^2 \, \left[  S_\mathrm{F_{th}}(\omega) + \vert \chi_\mathrm{eff,s} \vert^{-2} S_\mathrm{x_{n}}(\omega) \right] \label{eqn:Symp_S_y} \, .
\end{eqnarray}
By integrating equation (\ref{eqn:Symp_S_x}) we obtain the final steady state temperature $T_\mathrm{final}^\mathrm{sf}$ of combined feedback and sympathetic cooling:

\begin{equation}
T_\mathrm{final}^\mathrm{sf} \ = \ \frac{T_\mathrm{bath}}{\left(1+g_\mathrm{v} + g_\mathrm{s}  \right)} \, +  \frac{k_\mathrm{m} \omega_\mathrm{m} }{4 k_\mathrm{B} Q} \, \frac{g_\mathrm{v}^2}{\left(1+g_\mathrm{v} + g_\mathrm{s}  \right)} \, S_\mathrm{x_n}  \label{eqn:Tmode_sf} \, .
\end{equation}
\begin{figure}[!t]
    \includegraphics[width=1\textwidth]{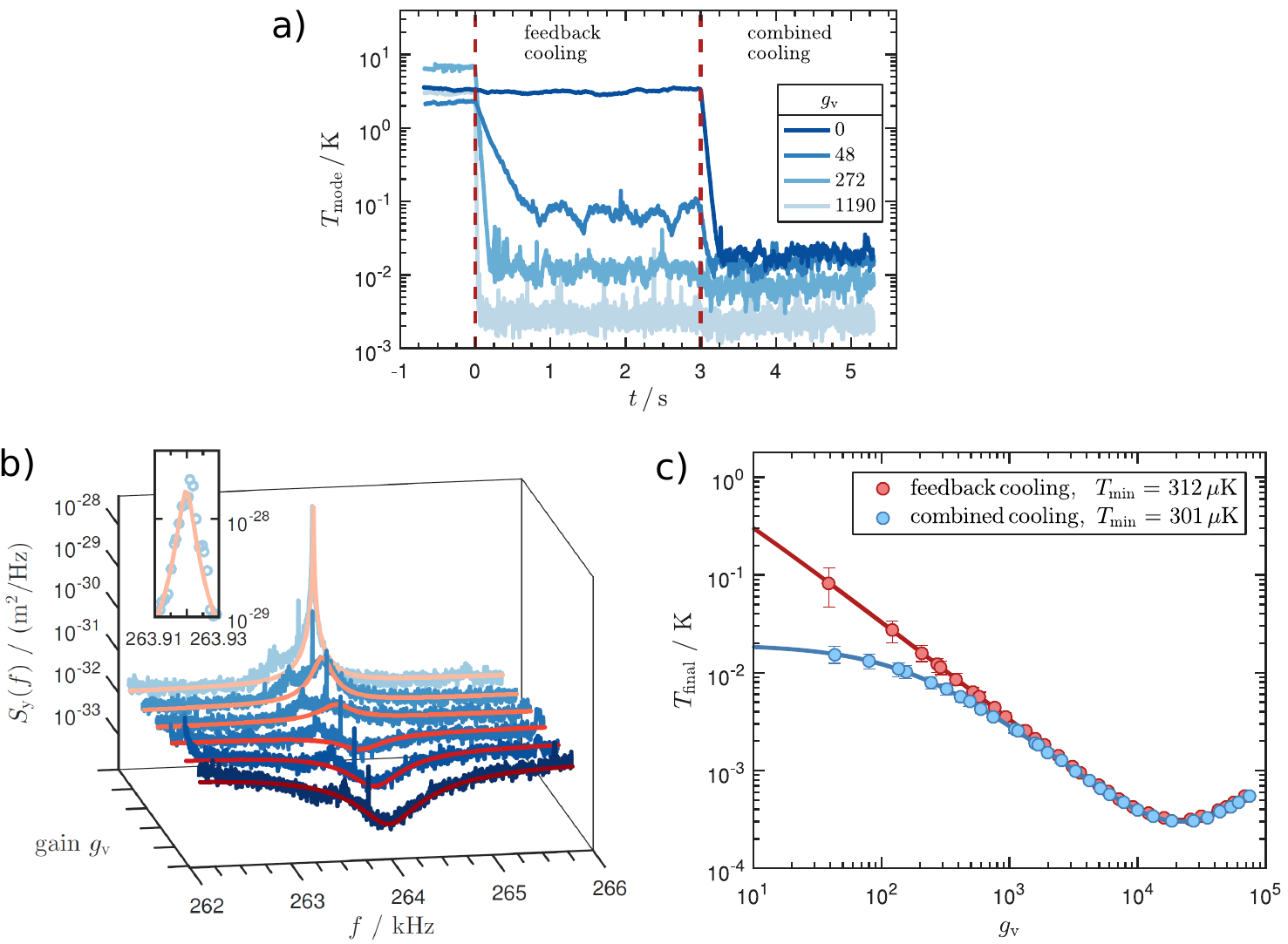}
     \caption{\label{fig:combined_results}\textbf{Combined sympathetic and feedback cooling.} 
     \textbf{a)} 	Zero-span traces of the membrane temperature $T_\mathrm{mode}(t)$ during the experimental sequence of combined cooling, as described in the text.
     \textbf{b)} 	In-loop PDS $S_\mathrm{y}(f)$ (blue) measured via zoom-FFT during the combined cooling slot and spectral fits to the data (red) using equation (\ref{eqn:Symp_S_y}) for different 			feedback gains $g_\mathrm{v}$. Inset: zoom into the peak area for $g_\mathrm{v}=0$.  
     \textbf{c)} 	Extracted mode temperatures $T_\mathrm{final}$ as a function of the fitted $g_\mathrm{v}$ from the spectra in b) and a fit to these temperatures according to equation 			(\ref{eqn:Tmode_sf}).}
\end{figure}
In order to validate this model we adjusted an experimental sequence of sympathetic MOT cooling and systematically altered the feedback gain $g_\mathrm{v}$, as shown in figure \ref{fig:combined_results}\,a. 
At $t=0$ we apply feedback and the membrane temperature $T_\mathrm{mode}$ reaches a new steady state $T_\mathrm{final}(g_\mathrm{v})$ given by equation (\ref{eqn:T_mode}). At $t= 3 \, \mathrm{s}$ sympathetic MOT cooling is applied in addition, as described in the previous section. 
The combined cooling leads to a new steady state temperature $T_\mathrm{final}^\mathrm{sf}(g_\mathrm{v}, g_\mathrm{s})$ given by equation (\ref{eqn:Tmode_sf}). 
Both steady state temperatures $T_\mathrm{final}$ and $T_\mathrm{final}^\mathrm{sf}$ were obtained by acquiring Zoom-FFT spectra $S^\mathrm{sf}_\mathrm{y}(\omega)$.
The spectral fits for combined cooling are shown in figure \ref{fig:combined_results}\,b using expression (\ref{eqn:Symp_S_y}) with $g_\mathrm{v}$ as the only free fitting parameter and $g_\mathrm{s} = 170$, which was extracted from the trace with $g_\mathrm{v} = 0$ in figure \ref{fig:combined_results}\,a. 
Similar to the pure feedback cooling spectra in figure \ref{fig:feedback}\,c, the combined cooling PSD $S^\mathrm{sf}_\mathrm{y}(\omega)$ can be fitted very well for all feedback gains $g_\mathrm{v}$. 
The additional noise peak left from the mechanical resonance is related to parasitic light from the coupling lattice beam entering the homodyne detection and it is also visible if no atoms are loaded into the MOT. 
We obtain $T_\mathrm{final}^\mathrm{sf}(g_\mathrm{v}, g_\mathrm{s})$ using equation (\ref{eqn:Tmode_sf}) and $g_\mathrm{v}$ from the spectral fits in figure \ref{fig:combined_results}\,b, as described above. 
The result is shown in figure \ref{fig:combined_results}\,c.
In good agreement with the behavior expected from equation (\ref{eqn:Tmode_sf}), we observe that for low feedback gains $g_\mathrm{v} < g_\mathrm{s}$ the feedback leads to significantly lower temperatures compared to pure sympathetic cooling. 
For large feedback gains $g_\mathrm{v} > g_\mathrm{s}$ the temperature of combined cooling is mostly determined by $g_\mathrm{v}$ and the effect of additional sympathetic cooling effect becomes very small.
Hence, the lowest achievable temperature of combined cooling at the optimal feedback gain differs only by a few percent from the temperature of pure feedback cooling.
However, the measurement shows that the hybrid coupling mechanism is compatible with feedback cooling of the membrane.
Note that for different experimental conditions, where sympathetic and optimal feedback gain are comparable, the combined cooling rate is significantly enhanced compared to each of the individual cooling methods as proposed in \cite{Bennett2014}.
It is worth noting that the minimum achievable membrane temperature $T_\mathrm{min}$ is 50\% larger than the value for pure feedback cooling. By directly comparing the feedback cooling performance with and without coupling lattice (and without atoms) we confirmed that this effect is caused by laser heating of the membrane through the coupling lattice. This can be significantly reduced by using mechanical oscillators with a lower resonance frequency, which allows for generating the resonance condition $\omega_\mathrm{a} = \omega_\mathrm{m}$ with much lower lattice powers $P_\mathrm{lat}$ at comparable lattice detunings $\Delta_\mathrm{lat}$.  

\section{Summary and experimental prospects}

In this paper, we have demonstrated active feedback cooling of the fundamental low-frequency mode of a Si$_3$N$_4$ membrane oscillator to a final occupation number of $\bar{n}_\mathrm{m} = 16 \pm 1$. Furthermore, we performed sympathetic cooling measurements of the membrane in our hybrid atomic-mechanical system \cite{Zhong2017}, which we used to determine the maximum hybrid cooperativity $C_\mathrm{hybrid} = 150 \pm 10$. By combining both cooling methods, we have shown that we are able to feedback cool the membrane into the quantum regime while it is coupled to the atoms. Moreover, our measurements represent the first realization of combined sympathetic and feedback cooling, which was proposed as a novel approach of ground state cooling in hybrid atomic-mechanical experiments \cite{Bennett2014}. 

Realsitic improvements of our setup will allow creating a  quantum hybrid system with the mechanical oscillator in the quantum ground state which is strongly coupled to an atomic ensemble in the quantum many-body ground state.
This can be achieved by replacing the simple square membrane with more elaborated types of mechanical oscillators \cite{Norte2016, Reinhardt2016}, which can be easily performed in our cryogenic optomechanical setup \cite{Zhong2017}. 
We expect feedback cooling into the quantum ground state for an oscillator with a ten times larger quality factor $Q_\mathrm{m}$ and a ten times lower oscillator mass m. 
This will also enhance the hybrid cooperativity $C_\mathrm{hybrid}$ by a factor of 100, as can be seen from equation (\ref{eqn:Gamma_sym}). 
Increasing the cavity finesse from the current value $\mathcal{F} \approx 160$ to an optimal value $\mathcal{F} \approx 850$ \cite{Vogell2013} in our hybrid system \cite{Bick2016} further increases $C_\mathrm{hybrid}$ by a factor of 30, since $C_\mathrm{hybrid} \sim \mathcal{F}^2$. 
With these improvements our hybrid system would satisfy the strong coupling condition $C_\mathrm{hybrid} > n_\mathrm{th}$, offering the possibilities of coherent quantum state transfer, teleportation and entanglement \cite{Hammerer2009, HammererZoller2009, Genes2011}.

\section{Acknowledgement}

We acknowledge experimental support in the early stage of the experiment from Christina Staarmann, Andreas Bick, Gotthold Fl\"{a}schner and Ortwin Hellmig.
Financial support by the DFG (grant No. BE 4793/2-1, SCHW 780/8-1, SE 717/9-1, WI 1277/29-1) is gratefully acknowledged.

\vspace{1cm}
\bibliography{library}

\begin{thebibliography}{10}
\newcommand{\enquote}[1]{``#1''}

\bibitem{Morton2011}
J.~J. Morton and B.~W. Lovett: \href
  {http://dx.doi.org/10.1146/annurev-conmatphys-062910-140514}
  {\enquote{{Hybrid Solid-State Qubits: The Powerful Role of Electron Spins}},
  }Annual Review of Condensed Matter Physics \textbf{2}~(1), 189--212 (2011),
  ISSN 1947-5454.

\bibitem{Degen2009}
C.~L. Degen, M.~Poggio, H.~J. Mamin, C.~T. Rettner and D.~Rugar: \href
  {http://dx.doi.org/10.1073/pnas.0812068106} {\enquote{{Nanoscale magnetic
  resonance imaging}}, }Proceedings of the National Academy of Sciences of the
  United States of America \textbf{106}~(5), 1313--7 (2009), ISSN 1091-6490.

\bibitem{Arcizet2011}
O.~Arcizet, V.~Jacques, A.~Siria, P.~Poncharal, P.~Vincent and S.~Seidelin:
  \href {http://dx.doi.org/10.1038/NPHYS2070} {\enquote{{A single NV defect
  coupled to a nanomechanical oscillator}}, }Nature Physics \textbf{7}~(11),
  879--883 (2011), ISSN 1745-2473.

\bibitem{Kolkowitz2012}
S.~Kolkowitz, A.~C. {Bleszynski Jayich}, Q.~P. Unterreithmeier, S.~D. Bennett,
  P.~Rabl, J.~G.~E. Harris and M.~D. Lukin: \href
  {http://dx.doi.org/10.1126/science.1216821} {\enquote{{Coherent Sensing of a
  Mechanical Resonator with a Single-Spin Qubit}}, }Science
  \textbf{335}~(6076), 1603--1606 (2012), ISSN 0036-8075.

\bibitem{Teissier2014}
J.~Teissier, A.~Barfuss, P.~Appel, E.~Neu and P.~Maletinsky: \href
  {http://dx.doi.org/10.1103/PhysRevLett.113.020503} {\enquote{{Strain Coupling
  of a Nitrogen-Vacancy Center Spin to a Diamond Mechanical Oscillator}},
  }Physical Review Letters \textbf{113}~(2), 020503 (2014), ISSN 0031-9007.

\bibitem{Ovartchaiyapong2014}
P.~Ovartchaiyapong, K.~W. Lee, B.~A. Myers and A.~C.~B. Jayich: \href
  {http://dx.doi.org/10.1038/ncomms5429} {\enquote{{Dynamic strain-mediated
  coupling of a single diamond spin to a mechanical resonator}}, }Nature
  Communications \textbf{5}, 1--6 (2014), ISSN 2041-1723.

\bibitem{Kagan2016}
C.~R. Kagan, E.~Lifshitz, E.~H. Sargent and D.~V. Talapin: \href
  {http://science.sciencemag.org/content/353/6302/aac5523.full}
  {\enquote{{Building devices from colloidal quantum dots}}, }Science
  \textbf{353}~(6302) (2016).

\bibitem{Yeo2013}
I.~Yeo, P.-L. de~Assis, A.~Gloppe, E.~Dupont-Ferrier, P.~Verlot, N.~S. Malik,
  E.~Dupuy, J.~Claudon, J.-M. G\'{e}rard, A.~Auff\`{e}ves, G.~Nogues,
  S.~Seidelin, J.-P. Poizat, O.~Arcizet and M.~Richard: \href
  {http://dx.doi.org/10.1038/nnano.2013.274} {\enquote{{Strain-mediated
  coupling in a quantum dot–mechanical oscillator hybrid system}}, }Nature
  Nanotechnology \textbf{9}~(2), 106--110 (2013), ISSN 1748-3387.

\bibitem{Montinaro2014}
M.~Montinaro, G.~W\"{u}st, M.~Munsch, Y.~Fontana, E.~Russo-Averchi, M.~Heiss,
  A.~{Fontcuberta I Morral}, R.~J. Warburton and M.~Poggio: \href
  {http://dx.doi.org/10.1021/nl501413t} {\enquote{{Quantum dot opto-mechanics
  in a fully self-assembled nanowire}}, }Nano Letters \textbf{14}~(8),
  4454--4460 (2014), ISSN 15306992.

\bibitem{Clarke2008}
J.~Clarke and F.~Wilhelm: \href {http://dx.doi.org/10.1038/nature07128}
  {\enquote{{Superconducting quantum bits}}, }Nature \textbf{453}~(7198),
  1031--1042 (2008), ISSN 0028-0836.

\bibitem{OConnell2010}
A.~D. O’Connell, M.~Hofheinz, M.~Ansmann, R.~C. Bialczak, M.~Lenander,
  E.~Lucero, M.~Neeley, D.~Sank, H.~Wang, M.~Weides, J.~Wenner, J.~M. Martinis
  and A.~N. Cleland: \href {http://dx.doi.org/10.1038/nature08967}
  {\enquote{{Quantum ground state and single-phonon control of a mechanical
  resonator}}, }Nature \textbf{464}~(7289), 697--703 (2010), ISSN 0028-0836.

\bibitem{Fink2016}
J.~M. Fink, M.~Kalaee, A.~Pitanti, R.~Norte, L.~Heinzle, M.~Davan\c{c}o,
  K.~Srinivasan and O.~Painter: \href {http://dx.doi.org/10.1038/ncomms12396}
  {\enquote{{Quantum electromechanics on silicon nitride nanomembranes}},
  }Nature Communications \textbf{7}, 12396 (2016), ISSN 2041-1723.

\bibitem{Palomaki2013}
T.~A. Palomaki, J.~W. Harlow, J.~D. Teufel, R.~W. Simmonds and K.~W. Lehnert:
  \href {http://dx.doi.org/10.1038/nature11915} {\enquote{{Coherent state
  transfer between itinerant microwave fields and a mechanical oscillator}},
  }Nature \textbf{495}~(7440), 210--214 (2013), ISSN 0028-0836.

\bibitem{Pirkkalainen2013}
J.-M. Pirkkalainen, S.~U. Cho, J.~Li, G.~S. Paraoanu, P.~J. Hakonen and M.~A.
  Sillanp\"{a}\"{a}: \href {http://dx.doi.org/10.1038/nature11821}
  {\enquote{{Hybrid circuit cavity quantum electrodynamics with a
  micromechanical resonator}}, }Nature \textbf{494}~(7436), 211--215 (2013),
  ISSN 0028-0836.

\bibitem{Wang2006}
Y.-J. Wang, M.~Eardley, S.~Knappe, J.~Moreland, L.~Hollberg and J.~Kitching:
  \href {http://dx.doi.org/10.1103/PhysRevLett.97.227602} {\enquote{{Magnetic
  Resonance in an Atomic Vapor Excited by a Mechanical Resonator}}, }Physical
  Review Letters \textbf{97}~(22), 227602 (2006), ISSN 0031-9007.

\bibitem{Hunger2010}
D.~Hunger, S.~Camerer, T.~W. H\"{a}nsch, D.~K\"{o}nig, J.~P. Kotthaus,
  J.~Reichel and P.~Treutlein: \href
  {http://dx.doi.org/10.1103/PhysRevLett.104.143002} {\enquote{{Resonant
  coupling of a bose-einstein condensate to a micromechanical oscillator}},
  }Physical Review Letters \textbf{104}~(14), 1--4 (2010), ISSN 00319007.

\bibitem{Camerer2011}
S.~Camerer, M.~Korppi, A.~J\"{o}ckel, D.~Hunger, T.~W. H\"{a}nsch and
  P.~Treutlein: \href {http://dx.doi.org/10.1103/PhysRevLett.107.223001}
  {\enquote{{Realization of an optomechanical interface between ultracold atoms
  and a membrane}}, }Physical Review Letters \textbf{107}~(22), 1--5 (2011),
  ISSN 00319007.

\bibitem{Montoya2015}
C.~Montoya, J.~Valencia, A.~A. Geraci, M.~Eardley, J.~Moreland, L.~Hollberg and
  J.~Kitching: \href {http://dx.doi.org/10.1103/PhysRevA.91.063835}
  {\enquote{{Resonant interaction of trapped cold atoms with a magnetic
  cantilever tip}}, }Physical Review A - Atomic, Molecular, and Optical Physics
  \textbf{91}~(6), 1--5 (2015), ISSN 10941622.

\bibitem{Moeller2017}
C.~B. M{\o}ller, R.~A. Thomas, G.~Vasilakis, E.~Zeuthen, Y.~Tsaturyan,
  M.~Balabas, K.~Jensen, A.~Schliesser, K.~Hammerer and E.~S. Polzik: \href
  {http://dx.doi.org/10.1038/nature22980} {\enquote{{Quantum
  back-Action-evading measurement of motion in a negative mass reference
  frame}}, }Nature \textbf{547}~(7662), 191--195 (2017), ISSN 14764687.

\bibitem{Bloch2012}
I.~Bloch, J.~Dalibard and S.~Nascimb\`{e}ne: \href
  {http://dx.doi.org/10.1038/nphys2259} {\enquote{{Quantum simulations with
  ultracold quantum gases}}, }Nature Physics \textbf{8}~(4), 267--276 (2012),
  ISSN 1745-2473.

\bibitem{Johanning2009}
M.~Johanning, A.~F. Varón and C.~Wunderlich: \href
  {http://stacks.iop.org/0953-4075/42/i=15/a=154009} {\enquote{Quantum
  simulations with cold trapped ions}, }Journal of Physics B: Atomic, Molecular
  and Optical Physics \textbf{42}~(15), 154009 (2009).

\bibitem{Chan2011}
J.~Chan, T.~P.~M. Alegre, A.~H. Safavi-Naeini, J.~T. Hill, A.~Krause,
  S.~Gr\"{o}blacher, M.~Aspelmeyer and O.~Painter: \href
  {http://dx.doi.org/10.1038/nature10461} {\enquote{{Laser cooling of a
  nanomechanical oscillator into its quantum ground state}}, }Nature
  \textbf{478}~(7367), 89--92 (2011), ISSN 1476-4687.

\bibitem{Teufel2011}
J.~D. Teufel, T.~Donner, D.~Li, J.~W. Harlow, M.~S. Allman, K.~Cicak, A.~J.
  Sirois, J.~D. Whittaker, K.~W. Lehnert and R.~W. Simmonds: \href
  {http://dx.doi.org/10.1038/nature10261} {\enquote{{Sideband cooling of
  micromechanical motion to the quantum ground state}}, }Nature
  \textbf{475}~(7356), 359--63 (2011), ISSN 1476-4687.

\bibitem{Peterson2016}
R.~Peterson, T.~Purdy, N.~Kampel, R.~Andrews, P.-L. Yu, K.~Lehnert and
  C.~Regal: \href {http://dx.doi.org/10.1103/PhysRevLett.116.063601}
  {\enquote{{Laser Cooling of a Micromechanical Membrane to the Quantum
  Backaction Limit}}, }Physical Review Letters \textbf{116}~(6), 063601 (2016),
  ISSN 0031-9007.

\bibitem{Zhong2017}
H.~Zhong, G.~Fl\"{a}schner, A.~Schwarz, R.~Wiesendanger, P.~Christoph,
  T.~Wagner, A.~Bick, C.~Staarmann, B.~Abeln, K.~Sengstock and C.~Becker: \href
  {http://dx.doi.org/10.1063/1.4976497} {\enquote{{A millikelvin all-fiber
  cavity optomechanical apparatus for merging with ultra-cold atoms in a hybrid
  quantum system}}, }Review of Scientific Instruments \textbf{88}~(2), 023115
  (2017), ISSN 0034-6748.

\bibitem{Vogell2013}
B.~Vogell, K.~Stannigel, P.~Zoller, K.~Hammerer, M.~T. Rakher, M.~Korppi,
  A.~J\"{o}ckel and P.~Treutlein: \href
  {http://dx.doi.org/10.1103/PhysRevA.87.023816} {\enquote{{Cavity-enhanced
  long-distance coupling of an atomic ensemble to a micromechanical membrane}},
  }Physical Review A - Atomic, Molecular, and Optical Physics \textbf{87}~(2),
  1--12 (2013), ISSN 10502947.

\bibitem{Joeckel2015}
A.~{J{\"o}ckel}, A.~{Faber}, T.~{Kampschulte}, M.~{Korppi}, M.~T. {Rakher} and
  P.~{Treutlein}: \href {http://dx.doi.org/10.1038/nnano.2014.278}
  {\enquote{{Sympathetic cooling of a membrane oscillator in a hybrid
  mechanical-atomic system}}, }Nature Nanotechnology \textbf{10}, 55--59
  (2015).

\bibitem{Bennett2014}
J.~S. Bennett, L.~S. Madsen, M.~Baker, H.~Rubinsztein-Dunlop and W.~P. Bowen:
  \href {http://dx.doi.org/10.1088/1367-2630/16/8/083036} {\enquote{{Coherent
  control and feedback cooling in a remotely coupled hybrid
  atom–optomechanical system}}, }New Journal of Physics \textbf{16}~(8),
  083036 (2014), ISSN 1367-2630.

\bibitem{Poggio2007}
M.~Poggio, C.~Degen, H.~Mamin and D.~Rugar: \href
  {http://dx.doi.org/10.1103/PhysRevLett.99.017201} {\enquote{{Feedback Cooling
  of a Cantilever’s Fundamental Mode below 5 mK}}, }Physical Review Letters
  \textbf{99}~(1), 1--4 (2007), ISSN 0031-9007.

\bibitem{Cohadon1999}
P.~Cohadon, A.~Heidmann and M.~Pinard: \href
  {http://dx.doi.org/10.1103/PhysRevLett.83.3174} {\enquote{{Cooling of a
  Mirror by Radiation Pressure}}, }Physical Review Letters \textbf{83}~(16),
  3174--3177 (1999), ISSN 0031-9007.

\bibitem{Wilson2014}
D.~J. Wilson, V.~Sudhir, N.~Piro, R.~Schilling, A.~Ghadimi and T.~J.
  Kippenberg: \href {http://dx.doi.org/10.1038/nature14672}
  {\enquote{{Measurement and control of a mechanical oscillator at its thermal
  decoherence rate}}, }Nature \textbf{524}~(7565) (2014), ISSN 0028-0836.

\bibitem{Courty2001}
J.~M. Courty, A.~Heidmaim and M.~Pinard: \href
  {http://dx.doi.org/10.1007/s100530170014} {\enquote{{Quantum limits of cold
  damping with optomechanical coupling}}, }European Physical Journal D
  \textbf{17}~(3), 399--408 (2001), ISSN 14346060.

\bibitem{Genes2008}
C.~Genes, D.~Vitali, P.~Tombesi, S.~Gigan and M.~Aspelmeyer: \href
  {http://dx.doi.org/10.1103/PhysRevA.77.033804} {\enquote{{Ground-state
  cooling of a micromechanical oscillator: Comparing cold damping and
  cavity-assisted cooling schemes}}, }Physical Review A - Atomic, Molecular,
  and Optical Physics \textbf{77}~(3), 1--10 (2008), ISSN 10502947.

\bibitem{Rossi2018}
M.~{Rossi}, D.~{Mason}, J.~{Chen}, Y.~{Tsaturyan} and A.~{Schliesser}: \href
  {http://arxiv.org/abs/1805.05087} {\enquote{{Measurement-based quantum
  control of mechanical motion}}, }ArXiv e-prints  (2018).

\bibitem{Bick2016}
A.~Bick, C.~Staarmann, P.~Christoph, O.~Hellmig, J.~Heinze, K.~Sengstock and
  C.~Becker: \href {http://dx.doi.org/10.1063/1.4939046} {\enquote{{The role of
  mode match in fiber cavities}}, }Review of Scientific Instruments \textbf{87}
  (2016).

\bibitem{Poot2011_Feedback}
M.~Poot, S.~Etaki, H.~Yamaguchi and H.~S.~J. {Van Der Zant}: \href
  {http://dx.doi.org/10.1063/1.3608148} {\enquote{{Discrete-time quadrature
  feedback cooling of a radio-frequency mechanical resonator}}, }Applied
  Physics Letters \textbf{99}~(1), 2--5 (2011), ISSN 00036951.

\bibitem{Poot2011}
M.~Poot and H.~S.~J. van~der Zant: \href
  {http://arxiv.org/abs/arXiv:1106.2060v2} {\enquote{{Mechanical systems in the
  quantum regime}}, }Physics Reports \textbf{511}~(5), 273--336 (2012).

\bibitem{Pinard2001}
M.~Pinard, P.~F. Cohadon, T.~Briant and A.~Heidmann: \href
  {http://dx.doi.org/10.1103/PhysRevA.63.013808} {\enquote{{Full mechanical
  characterization of a cold damped mirror}}, }Physical Review A - Atomic,
  Molecular, and Optical Physics \textbf{63}~(1), 1--12 (2001), ISSN 10502947.

\bibitem{Aspelmeyer2014}
M.~Aspelmeyer, T.~J. Kippenberg and F.~Marquardt: \href
  {http://dx.doi.org/10.1103/RevModPhys.86.1391} {\enquote{{Cavity
  optomechanics}}, }Reviews of Modern Physics \textbf{86}~(4), 1391--1452
  (2014).

\bibitem{Norte2016}
R.~Norte, J.~Moura and S.~Gr\"{o}blacher: \href
  {http://journals.aps.org/prl/abstract/10.1103/PhysRevLett.116.147202}
  {\enquote{{Mechanical Resonators for Quantum Optomechanics Experiments at
  Room Temperature}}, }Physical Review Letters \textbf{116}~(14), 147202
  (2016), ISSN 0031-9007.

\bibitem{Reinhardt2016}
C.~Reinhardt, T.~M\"{u}ller, A.~Bourassa and J.~C. Sankey: \href
  {http://journals.aps.org/prx/abstract/10.1103/PhysRevX.6.021001}
  {\enquote{{Ultralow-Noise SiN Trampoline Resonators for Sensing and
  Optomechanics}}, }Physical Review X \textbf{6}~(2), 021001 (2016), ISSN
  2160-3308.

\bibitem{Raithel1998}
G.~Raithel, W.~D. Phillips and S.~L. Rolston: \href
  {http://dx.doi.org/10.1103/PhysRevLett.81.3615} {\enquote{{Collapse and
  revival of wave packets in optical lattices}}, }Physical Review Letters
  \textbf{81}~(17), 3615--3618 (1998).

\bibitem{Ovchinnikov1999}
Y.~B. Ovchinnikov, J.~H. M\"{u}ller, M.~R. Doery, E.~J.~D. Vredenbregt,
  K.~Helmerson, S.~L. Rolston and W.~D. Phillips: \href
  {http://dx.doi.org/10.1103/PhysRevLett.83.284} {\enquote{{Diffraction of a
  Released Bose-Einstein Condensate by a Pulsed Standing Light Wave}},
  }Physical Review Letters \textbf{83}~(2), 284--287 (1999), ISSN 0031-9007.

\bibitem{Aline2017}
A.~{Vochezer}, T.~{Kampschulte}, K.~{Hammerer} and P.~{Treutlein}: \href
  {http://arxiv.org/abs/1705.10098} {\enquote{{Light-mediated collective atomic
  motion in a hybrid atom-optomechanical system}}, }ArXiv e-prints  (2017).

\bibitem{Asboth2007}
J.~K. Asb\'{o}th, H.~Ritsch and P.~Domokos: \href
  {http://dx.doi.org/10.1103/PhysRevLett.98.203008} {\enquote{{Collective
  Excitations and Instability of an Optical Lattice due to Unbalanced
  Pumping}}, }Physical Review Letters \textbf{98}~(20), 203008 (2007), ISSN
  0031-9007.

\bibitem{Asboth2008}
J.~K. Asb\'{o}th, H.~Ritsch and P.~Domokos: \href
  {http://dx.doi.org/10.1103/PhysRevA.77.063424} {\enquote{{Optomechanical
  coupling in a one-dimensional optical lattice}}, }Physical Review A
  \textbf{77}~(6), 063424 (2008), ISSN 1050-2947.

\bibitem{Hammerer2009}
K.~Hammerer, M.~Wallquist, C.~Genes, M.~Ludwig, F.~Marquardt, P.~Treutlein,
  P.~Zoller, J.~Ye and H.~J. Kimble: \href
  {http://dx.doi.org/10.1103/PhysRevLett.103.063005} {\enquote{{Strong coupling
  of a mechanical oscillator and a single atom}}, }Physical Review Letters
  \textbf{103}~(6), 2--5 (2009), ISSN 00319007.

\bibitem{HammererZoller2009}
K.~Hammerer, M.~Aspelmeyer, E.~S. Polzik and P.~Zoller: \href
  {http://dx.doi.org/10.1103/PhysRevLett.102.020501} {\enquote{{Establishing
  Einstein-Poldosky-Rosen Channels between Nanomechanics and Atomic
  Ensembles}}, }Physical Review Letters \textbf{102}~(2), 020501 (2009), ISSN
  0031-9007.

\bibitem{Genes2011}
C.~Genes, H.~Ritsch, M.~Drewsen and A.~Dantan: \href
  {http://dx.doi.org/10.1103/PhysRevA.84.051801} {\enquote{{Atom-membrane
  cooling and entanglement using cavity electromagnetically induced
  transparency}}, }Physical Review A \textbf{84}~(5), 051801 (2011), ISSN
  1050-2947.

\end{thebibliography}

\end{document}